\documentclass{article}
\usepackage{algo}

\newcommand{\RRonly}[1]{}
\newcommand{\conf}[1]{}
\newcommand{\confsubmit}[1]{}
\newcommand{\journal}[1]{#1}

\journal{
}

\conf{
\usepackage{stoc98}
\usepackage{times}
\setlength{\textheight}{22cm}
\setlength{\columnsep}{0.33in}
\setlength{\textwidth}{7in}
\setlength{\topmargin}{0.25in}
\setlength{\headheight}{0.0in}
\setlength{\headsep}{0.0in}
\setlength{\oddsidemargin}{-.19in}
\setlength{\parindent}{1pc}
\setlength{\parskip}{0.1in}
\sloppy
\pagestyle{empty}
}

\RRonly{\usepackage{RR}}

\confsubmit{
\voffset=-2cm
\hoffset=-1cm
\textwidth              16cm
\textheight             23cm

}

\newcommand{\ABSTRACT}{
This paper presents how the space of spheres and shelling
may be used to delete a point from a
$d$-dimensional triangulation efficiently.
In dimension two, if $k$ is the degree of the deleted vertex,
the complexity is $O(k\log k)$,
but we  notice that this number only applies
to low cost operations, while time consuming computations are
 only done a linear number of times. 

This algorithm may be viewed as a variation
of Heller's algorithm \cite{h-taatm-90,m-smdnt-93},
which is popular in the geographic information system
community. Unfortunately, Heller algorithm is false,
as explained in this paper.
}

\begin{document}

\journal{
\title{On Deletion in  Delaunay Triangulations.\thanks{
        This work was partially supported by ESPRIT LTR 21957 (CGAL)}}
\author{Olivier Devillers\thanks{
       INRIA, BP93, 06902 Sophia Antipolis, France.
        Olivier.Devillers@sophia.inria.fr.
       \vspace*{2.5cm}
}}
\date{}
\maketitle
\thispagestyle{empty}
\abstract{\ABSTRACT}
}

\section{Introduction}

The computation of the Delaunay triangulation of
a set {\cal S} of $n$ points in the plane is one of the classical
problems of computational geometry.

Many structures and algorithms
have been proposed in the past to compute
Delaunay triangulations.
Some of these algorithms have the two following properties:
they are incremental and they do not used complicated data structures
in addition to the triangulation itself.
Among these algorithms, let us mention the historical algorithm
of Green and Sibson \cite{gs-cdtp-78}, or some other variants
\cite{msz-frplw-96,bd-irgo-95,prisme-3298i,dlm,l-tdam-97}.
All perform a walk in the triangulation to accelerate point location.

The advantage of that category of incremental  Delaunay algorithms
is that they may easily be turned into fully dynamic Delaunay
algorithms. Since there is no complicated data structure
for point location, the deletion of a point is reduced to the
deletion in the triangulation itself.

{\bf Definition and notations}

Given  a set $\cal S$ of points in $d$-dimensional space,
 $\cal DT(S)$, the Delaunay triangulation of $\cal S$
is defined by the following property:
{\em $d+1$ points of $\cal S$ are the vertices of a Delaunay
simplex if and only if the sphere passing through these
points does not contain another point of $\cal S$ in its interior}
(see Figure \ref{Example} for a Delaunay triangulation in
two dimensions).

Given the Delaunay triangulation $\cal DT(S)$
 and
a vertex $p$ in $\cal DT(S)$,
we address the problem of finding ${\cal DT(S}\setminus\{p\})$.

In two dimensions, the natural parameter to evaluate the
complexity of this problem is the degree $k$ of $p$
in  $\cal DT(S)$, since the deletion of $p$ means that
$k$ triangles must be removed from the triangulation and
$k-2$ new triangles must be created to fill this hole.
In the worst case, $k$ may be $|{\cal S}|$, but if $p$ is
chosen randomly in $\cal S$, then it is well known that the
expected value of $k$ is 6, without any assumption
on the point distribution.

In higher dimensions, the number of simplices incident to $p$
is not directly related to the number of simplices created
to fill the hole. We will let $f$ denote the sum of these two numbers.
In the worst case, the whole Delaunay triangulation
may be affected, and $f=O(n^{\lfloor \frac{d+1}{2}\rfloor})$.
This distribution does not reflect practical configurations,
and a constant value of $f$ is more likely in practice.
For a uniform point distribution, the expected value of $f$
may be shown to be constant.
In three dimensions, the expected number of
deleted tetrahedron for Poisson distribution
is $\frac{96}{35}\pi ^2 \simeq 27$ \cite{obs-stcav-92}.

Thus, even if we do
not want to neglect the possible case of a big value for $k$,
we have to keep in mind that a good algorithm must perform well
on small values of $k$.

{\bf Previous related work}

Classical Computational Geometry has already addressed the
problem of deleting points from Delaunay triangulations.
This can be done with optimal asymptotic complexity
$O(k)$ in 2 dimensions \cite{c-bvdcp-86,agss-ltacv-89}.
But these algorithms are a little bit too intricate
and the big $O$ hides  too important a constant for them to
be good algorithms for reasonable values of $k$
(we give this constant in Section \ref{compare-Chew}
for Chew's algorithm, Aggarwal et al's algorithm
is even more complicated).

Practitioners often prefer algorithmic simplicity
to theoretical optimality, and  favour a 
simple suboptimal $O(k^2)$, implementation of the deletion algorithm.
This may be achieved, for example, by flipping, to reduce
the degree of the deleted vertex to 3, and flipping again to restore
the Delaunay property. Another simple algorithm
 consists in finding the Delaunay triangle incident to an edge
of the hole in $O(k)$ time, which also yields an $O(k^2)$ time algorithm.

A very simple $O(k\log k)$ solution was suggested
by Heller \cite{h-taatm-90,m-smdnt-93},
in which successive ears are filled in turn. In that way,
during the algorithm we always have
 a simple polygon of decreasing size to triangulate.
Unfortunately, this solution
is wrong, but we will show in this paper how to correct it.

{\bf Overview}

In this paper, we provide a very simple and efficient $O(k\log k)$
algorithm to delete a vertex in a planar Delaunay triangulation
based on shelling \cite{bm-sdcs-71,s-chdch-86} and duality
\cite{dmt-ssgtu-92i,p-gcc-70}.
We also discuss the effective complexity of this algorithm
and of a few others for small values of $k$.
We will study the different kinds of geometric
predicates necessary for these algorithms.

This algorithm generalizes well in higher dimensions: its time complexity
becomes $O(f\log f)$, where $f$ is the number of tetrahedra created,
and it  also generalizes to regular triangulations (power diagrams).

\section{Two dimensional algorithm\label{algo2d}}
\begin{figure} \begin{center} \def\IPEfile{Example.ipe}\input{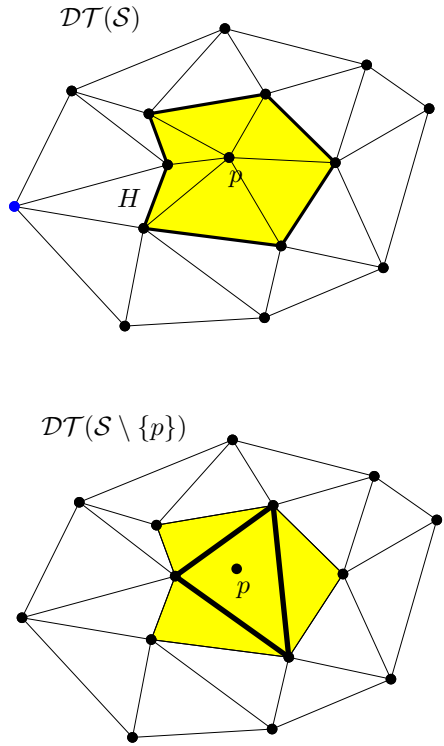} 
\caption{\label{Example}Deletion of a vertex.}
\end{center} \end{figure}

A deletion algorithm has to remove all triangles incident
to $p$ and retriangulate ``Delaunay-wise'' the star-shaped polygon
$H=\{q_0,q_1,\ldots,q_{k-1},q_k=q_0\}$ created by
these removals (see Figure \ref{Example}).

\subsection{Ears, and a wrong algorithm}

\begin{figure} \begin{center} \def\IPEfile{Radius-ear.ipe}\input{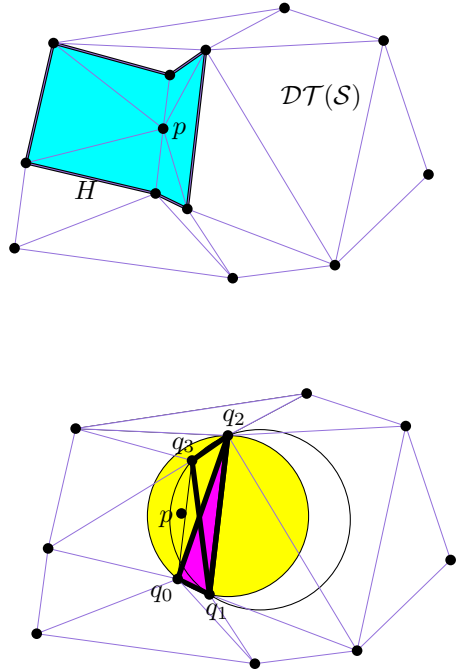} 
\caption{\label{Radius-ear}The smallest potential ear may
not belong to the final Delaunay triangulation.
The shaded triangle is the ear with smallest circumradius, but its
associated circle contains
$q_3$ and thus is not a Delaunay ear.}
\end{center} \end{figure}

We first define what an ear of a polygon is.
Three consecutive  vertices $q_iq_{i+1}q_{i+2}$
along $H$'s boundary
 are said to form an ear of $H$
if the line segment $q_iq_{i+2}$ is inside $H$
and does not cross its boundary.
An ear of $H$ is said to be Delaunay, if the circle
 through $q_i$ $q_{i+1}$ and $q_{i+2}$ does not contain
any other vertices of $H$ in its interior.
 Heller \cite{h-taatm-90} (also cited by Midtb{\o} \cite{m-smdnt-93})
claimed (without proof) that
among all the potential ears $q_iq_{i+1}q_{i+2}$
of $H$, the one having the  circumcircle with smallest radius
is a Delaunay ear.
This claim is false, as illustrated by Figure \ref{Radius-ear}:
on the left handside 
are shown the Delaunay triangulation
$\cal DT(S)$ and the hole $H$ to be retriangulated;
on the right handside are shown two potential ears
$q_0q_1q_2$ and $q_1q_2q_3$.
 $q_0q_1q_2$ has the  smallest circumcircle among all ears
of $H$, but it contains $q_3$, which invalidates Heller's claim.
Heller's mistake  is to assume that when
we deform a circle through $q_1q_2$ to maximise its portion
 inside $H$, the radius
 increases, but this is true only if the center of the circle is inside $H$,
which is not the case in Figure  \ref{Radius-ear}.

In fact, the idea of finding an ear belonging to the
Delaunay triangulation  works, but with another criterion,
as explained below.

\subsection{Delaunay and convex hull}

There exists a well known duality between Delaunay triangulations
in dimension $d$ and convex hulls in dimension $d+1$
\cite{a-pdpaa-87,dmt-ssgtu-92i}.
If we associate to a point
$p=(x,y)\in\cal S$ a point $p^{\star}=(x,y,x^2+y^2)$ on the paraboloid $\Pi$
of equation $z=x^2+y^2$,
the Delaunay triangulation of $\cal S$ is the projection
of the convex hull of the 3D points.
The reason is that for
 $p,q,r,s\in \cal S$,
$p$ is inside the circle $C_{qrs}$ through $qrs$ if and only if
$p^{\star}$ is below the plane $P_{qrs}$ through $q^{\star}r^{\star}s^{\star}$
($\Pi\cap P_{qrs}$ projects onto  circle  $C_{qrs}$).
We even have the equality between the power of $p$ with respect
to $C_{qrs}$ and the signed vertical distance between $p^{\star}$
and $P_{qrs}$.\footnote{
If $C$ is a circle of center $x$ and radius $r$, $p$ is a point
and $l$ is a line through $p$ intersecting $C$ in $t$ and $u$, then
$power(p,C)= |xp|^2-r^2 = \overline{pt} \overline{pu}$
where $\overline{yz}$ is the signed length of $yz$.
$power(p,C)$ is zero on $C$ boundary, negative inside and
positive outside. When $C$ is given by three points, the power
may be known without computing the circumradius, as explained in Section
\ref{power formula}. Power is negative inside the circle and
positive outside.
}

\subsection{Shelling}

Convex hulls may be computed by the shelling algorithm
\cite{s-chdch-86}. The  shelling \cite{bm-sdcs-71} of
a convex polyhedron $P$ is the enumeration of its faces in some
appropriate order.
Imagine  an observer is moving along
a line $l$ going through the polyhedron, starting at the intersection
of $P$ and $l$.
At the starting position, the observer can  only see one face of $P$
(the face intersecting its trajectory) and when she moves
away from $P$
she discovers other faces one by one;
at infinity, the observer sees ``half'' of $P$.
Then the observer
turns round at infinity on the opposite side of $l$
(where she sees the other half of $P$),
and then moves on $l$, enumerating
 the faces of $P$ when they disappear from
her view.
This order is called the shelling order of $P$ with respect to $l$;
it has the property that the set of enumerated faces remains simply connected
during the enumeration.
Seidel's algorithm for convex hull reports the faces of the convex hulls
in that order, by
maintaining a priority queue of potential new faces.
These new faces are of two kinds depending on whether
 the next face contains a new
vertex, or else connects already visible vertices.

\subsection{Deletion in Delaunay}

Let us now  use the idea of finding Delaunay ears to
retriangulate the hole created by the deletion of a point $p$
in $\cal DT(S)$.
Using duality with convex hulls, the problem is transformed into filling
the hole in the convex hull created by the deletion of $p^{\star}$.
Then we may note that if we use  shelling order with respect to
the vertical line through $p^{\star}$, an observer (going up) reaching
$p^{\star}$ sees the boundary of this hole exactly;
thus, the end of the shelling procedure
corresponds exactly to the triangulation
of the hole.
This partial shelling is easier to implement than Seidel's algorithm,
since all the vertices are already  visible and thus
only one kind of potential new faces has to be found.
The already noted correspondence between vertical distance and power
yields the following lemma.

\begin{quote} {\bf Lemma:} \em
Consider a polygon $H=\{q_0,q_1,\ldots,q_{k-1},q_k=q_0\}$
and a point $p$ such that the edges of $q_iq_{i+1}$ belongs
to the Delaunay triangulation of
$\{q_0,q_1,\ldots,q_{k-1},p\}$.
If $|power(p,circle(q_i,q_{i+1},q_{i+2}))|$ is minimal,
then $q_iq_{i+2}$ is an edge of
the Delaunay triangulation of
$\{q_0,q_1,\ldots,q_{k-1}\}$
\end{quote}

Thus the deletion of a point may be implemented in
a simple way, by maintaining a structure to store the ears.
The ears are naturally ordered along the boundary of the hole,
 each with its priority.
This structure must support the following operations~:
find next and previous ear according to counterclockwise order along $H$'s
boundary,
delete ear with minimum priority
and update the priority of an ear.
This structure may be implemented with any dictionary structure
augmented by next and previous pointers.

\begin{algorithm}{Delete}[{\cal DT(S)},p)]{}
  Let $q_0q_1\ldots q_{k-1}$ be the vertices incident to $p$
    in $\cal DT(S)$ in ccw order around $p$;\\
  Let $Q$ be a priority queue;\\
  \qfor $i = 0$ {\bf to} $k-1$\\
  \qdo $ear$ \qlet $q_iq_{i+1}q_{i+2}$;\\
       \qif counterclockwise($q_iq_{i+1}q_{i+2}$) \\
                 \qthen $p$ \qlet $\infty$;   \hfill{\em //not an ear}\\
             \qelse $p$ \qlet $-power(p,ear)$;\hfill{\em//inside circle power$<0$}
        \qfi\\
         $Q$.insert($p$,$ear$); \hfill{\em //insert(priority,key)}
  \qrof\\
  \qwhile $Q$.size()$>3$\\
  \qdo $ear$ \qlet $Q$.minimum(); \\
       create triangle $ear$ and link it to its two existing neighbors;\\
       $ear0$ \qlet $ear$.previous;\\
       $ear1$ \qlet $ear$.next;\\
       $ear0$.vertex(2) \qlet $ear$.vertex(2); $ear0$.next \qlet $ear1$;\\
       $ear1$.vertex(0) \qlet $ear$.vertex(0); $ear1$.previous \qlet $ear2$;\\
       $Q$.delete($ear$);\\
       $Q$.modify-priority($ear0$);\\
       $Q$.modify-priority($ear1$);
  \qelihw\\
  $ear$ \qlet $Q$.minimum();\hfill{\em //the three last ears are identical}\\
  create triangle $ear$ and link it to its three existing neighbors;
 \qend
\end{algorithm} 

{\bf Higher dimensions}
The generalization to $d$ dimensions is easy.
The boundary of the region to retriangulate is a simple polyhedron $H$,
and the ears are simplices formed by the vertices
of two incident facets of $H$. The difference is that the same simplex
may correspond to $O(d^2)$  pairs of incident facets, and that the creation
of an ear may modify  $O(d^2)$ other ears in the priority queue.

\section{2D Analysis\label{analysis}}
\subsection{Power computation\label{power formula}}
An analytical expression of the power of $p$ with respect to $q_0q_1q_2$ is
\begin{small}
\vbox{
\begin{eqnarray*}
\lefteqn{
power(p,circle(q_0,q_1,q_2))}\\
&=&
\frac
{
 \left|\begin{array}{cccc}
 x_{q_0} & x_{q_1} & x_{q_2} & x_p \\
 y_{q_0} & y_{q_1} & y_{q_2} & y_p \\
 x_{q_0}^2+y_{q_0}^2 & x_{q_1}^2+y_{q_1}^2 & x_{q_2}^2+y_{q_2}^2 & x_p^2+y_p^2\\
 1 & 1 & 1& 1 \end{array} \right|
}{
 \left| \begin{array}{ccc}
 x_{q_0} & x_{q_1} & x_{q_2} \\
 y_{q_0} & y_{q_1} & y_{q_2} \\
 1 & 1& 1 \end{array} \right|
}
\end{eqnarray*}
}
\end{small} 
The $3\times 3$ determinant is the orientation test of $q_0q_1q_2$,
and the $4\times 4$ determinant the incircle test of $p$ with respect
to $q_0q_1q_2$.
First note that if $q_0q_1q_2$ has the wrong orientation,
then it is not an ear and thus the $4\times 4$ determinant does
not need to be computed, and also that the orientation test is a minor
of the incircle test, and thus the power computation just requires one
division in addition to the usual incircle test.

Using a dynamic programming development of the determinant,
the power computation requires
14 additions, 15 multiplications and one division.

Finally, note that
 if only $q_2$ changes, we  do not need to recompute everything.
In fact, we can do it with
6 additions, 7 multiplications and one division.


\subsection{Complexity}

The theoretical asymptotic complexity of
this  algorithm is clearly $O(k\log k)$,
but this complexity  only concerns the
management of the priority queue, which involves relatively cheap
operations (pointer manipulations and comparisons of computed powers).

Since the most expensive geometric operations are the power computations,
we shall count the number of such operations exactly.
The initial size of the priority queue is  $k$,
and thus its initialization requires at most $k$ power computations.
Each ear creation implies the modification of two other ears, and thus two
powers must be recomputed,
and the deletion is completed when the size
of the queue is 3;
thus the total number of power computations is $k+2(k-4)=3k-8$.

It is possible, as noticed above, to update the power of $p$ with respect
to $q_iq_{i+1}q_{i+2}$ when ear $q_{i+1}q_{i+2}q_{i+3}$
is processed.
In the new polygon $H\setminus \{q_{i+2}\}$, $q_iq_{i+1}q_{i+3}$
is an ear and the power of $p$ with respect
to $q_iq_{i+1}q_{i+3}$ may be obtained by updating the power
of $p$ with respect to  $q_iq_{i+1}q_{i+2}$ at a cheaper cost.
Then the total number of computations becomes
$2k-4$ power computations and $k-4$ power updates.


\subsection{Robustness issues and degeneracies}

The algorithm presented above does not address robustness issues.
If floating point arithmetic is used to perform power computations,
the results are rounded and their comparisons could be evaluated
erroneously.
We can first observe that the deletion algorithm will terminate
even with incorrect arithmetic: it fill ears in turn and thus constructs
a topological triangulation, which may be non Delaunay, or even have
a non-planar embedding. But, even if producing a non-exact Delaunay
triangulation may be acceptable for the deletion algorithm, it is
 unacceptable for many insertion algorithms which are not capable
of processing non-exact triangulations.

The usual way to solve robustness issues consists in using exact
arithmetic. To ensure good performance, we can use arithmetic
filters to use exact computations in power comparisons only
in difficult cases where the two powers are close.
Exact computations are then used to take the right decision.
This approach of filtering out easy cases has been proven
efficient on the orientation and incircle tests \cite{prisme-2971a,prisme-3528i}.
Degeneracies can be solved using perturbation techniques
\cite{s-nmpgc-98,prisme-3316t}.

\section{Alternative methods and practical results}

\subsection{Alternative methods in  two dimensions}

\paragraph{Diagonal flipping}
One of the interests of a method using diagonal flipping is
that it does not introduce new geometric predicates:
it only  uses incircle tests, and thus has lower degree
and generalizes easily to various metrics.
However such a method may require $O(k^2)$ incircle tests
in the worst case, and is not so simple to code efficiently.
A good implementation of a flipping method \cite{jack}
will retriangulate $H$ by basically linking all $q_i$ to
$q_0$ and flipping the edges turning around $q_0$.
In the worst case, such a method may use
$(k-3)+(k-4)+\ldots+1=\frac{(k-2)(k-3)}{2}$
incircle tests.

As this flipping method is not so easy to code, a simpler
solution may consist in maintaining a queue of edges to be tested
for potential flip. Each time the diagonal of a quadrilateral
is flipped, the four edges of the quadrilateral are inserted in the queue.
This simpler algorithm will make much more incircle tests,
since the flip are not performed in some relevant order as above.

\paragraph{Edge completion}
A second method consists in finding $q_i$ such that
$q_0q_1q_i$ is a Delaunay triangle, which may be done
in $k-3$ incircle tests, and triangulating recursively the two holes.
In the worst case, the number of incircle tests is
exactly the same that in the flipping method.

\paragraph{Randomized algorithm\label{compare-Chew}}
In Chew \cite{c-bvdcp-86} randomized algorithm, each point to
reinsert requires an expected number of 5 incircle tests,
which yields a total of $5k+O(1)$.

\paragraph{On alternative methods}
For small value of $k<9$, flipping or edge completion requires less
incircle tests  computations, but the simplicity of our 
``ear-queue'' algorithm
and its good performance for $k\geq 9$ make it a very good candidate
for Delaunay vertex deletion.
Nevertheless, it can be interesting to treat as special cases
some small $k$ values such as $k=4$ and $k=5$.

\subsection{Higher dimensions}
 In higher dimensions, flipping, edge completion and shelling
algorithms generalize but things became more difficult.
The flipping must be done in a higher dimension, which makes  it
more intricate to implement \cite{es-itfwr-96}.
Edge completion transforms into facet completion, and 
must deal with the triangulation of non simply connected polyhedra.
Shelling is the easiest method to generalize.
Furthermore, the increase in the average value of $k$ with the dimension
reinforces its advantage over alternative candidates in higher dimensions.

\subsection{Experimental results}

\paragraph{Code and data}

This algorithm was implemented within the author's simple
hierarchical structure \cite{prisme-3298i}.

Robustness issues are solved using 24 bits integers
to store points coordinates.
Geometric predicates are computed with approximate
arithmetic and the exactness of the result
can be ensured by static
and semi-static filters\footnote{
Filters can be classified in static, semi-static and dynamic 
\cite{bbp-iayed-98scg}:\\
{\bf static:} error bound is determined at compile time,\\
{\bf dynamic:} error bound is determined at run time, usually
with an error computation for all intermediate results,\\
{\bf semi-static:} error bound is mainly determined at compile time,
with addition of very few computation at run time.\\
}.
The  filters failure are backed up by exact computations.

According to paragraphs above,
we have coded several versions of the deletion procedure.
\begin{itemize}
\item {\bf ear3.}
The basic version using the ear queue  explained in this paper.
\item {\bf ear5.}
 A variant processing  the ears' queue while its size is greater than five.
 When the region to triangulate is a pentagon, a
specific method using two or three incircle tests is used .
Degree three, four and five vertices are removed by a specific algorithms.
\item {\bf flip.}
The flipping-based method briefly described above.
\item $d_{limit}$.
A {\bf mixed} method using {\bf ear5}  if the degree of the removed point
is $\geq d_{limit}$ and the {\bf flip} method  otherwise.
\end{itemize}

\paragraph{Results}

\begin{figure} \begin{center} \def\IPEfile{Random.ipe}\input{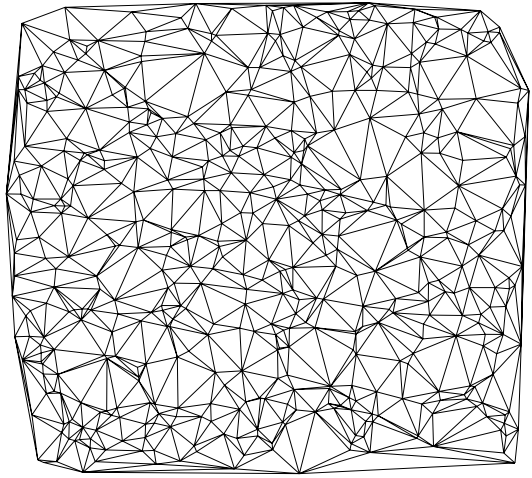} 
\caption{\label{Random}Delaunay triangulation of 
random points in a square.}
\end{center} \end{figure}

\begin{figure} \begin{center} \def\IPEfile{Stats.ipe}\input{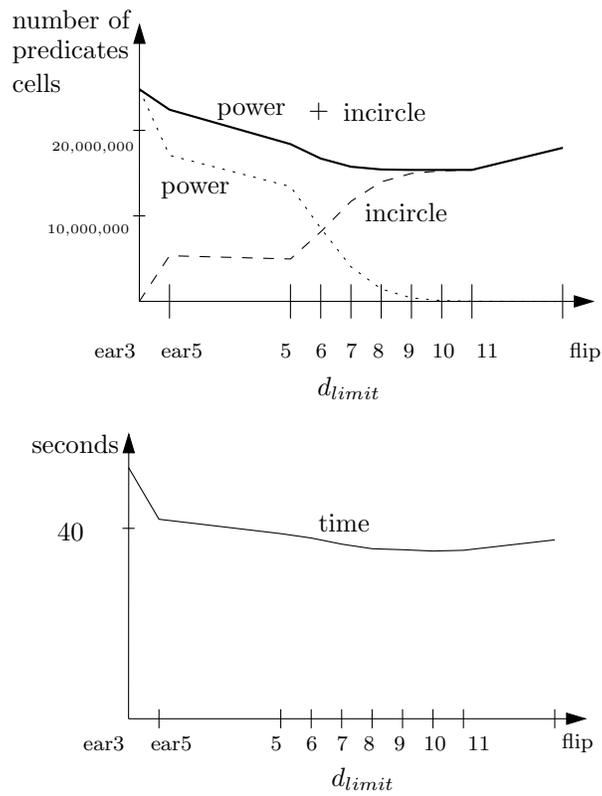} 
\caption{\label{Benchs}Deletion time for two millions of
random points in a square for different deletion methods.
See paragraph ``Results'' for details.}
\end{center} \end{figure}

The code was tested on a 2,000,000 points set uniformly
 distributed in a square (Figure \ref{Random}).
The Delaunay triangulation of the points is first computed
(in 104 seconds).
Next the points are  removed in a random order.
We tried the methods ear3, ear5, flip and the mixed method for 
$5\leq d_{limit} \leq 11$.
Figure \ref{Benchs} provides the whole deletion time,
the number of incircle predicate evaluations and the number of
power computations.

These experiments have been done
 on  a Sun Ultra10 300MHz workstation 256Mo main memory.
  The code is written in C++ and compiled with AT-T's compiler with
  optimizing options.
   Times were obtained with the {\tt clock} command
  and are given in seconds and are only for the deletion phase.

As infered from the theory, the performance is optimal
when $d_{limit}$ is about 9.
The mixed method therefore reduces the complexity
of the deletion of high degree vertices from $O(k^2)$ to $O(k\log k)$.

\newpage

\begin{figure}[h] \begin{center} \def\IPEfile{Lower-bound.ipe}\input{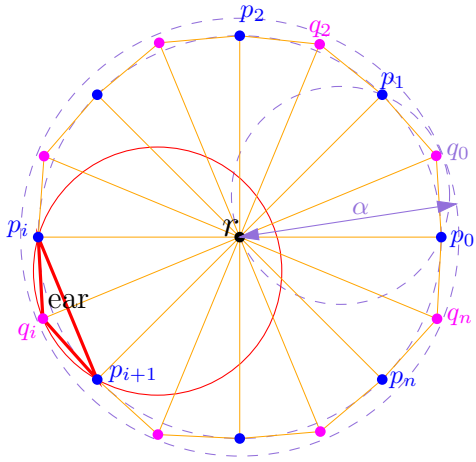} 
\caption{\label{Lower-bound} For $\Omega(k\log k)$ lower bound.}
\end{center} \end{figure}

\section{Lower bound}

The ear-queue method, i.e. the construction of the ears in the right order,
has a $\Omega(k\log k)$ lower bound. 
Consider the origin $r$ and  points
$p_i, 0\leq i \leq n$ and $q_i, 0\leq i \leq n$ so that 
angles $p_irq_i=q_irp_{i+1}=\frac{\pi}{n}$ and distances
$p_ir=1$ and $q_ir=x_i, 1<x_i<\alpha$.
$\alpha$ is chosen so that the Delaunay triangulation of
${\cal S}=\{r,p_0\ldots p_n,q_0\ldots q_n\}$
links $r$ to all other points (Figure \ref{Lower-bound}).
 When deleting $r$,
all $p_iq_ip_{i+1}$ are Delaunay ears, but finding the right order
on this ears, which is not necessary to find the new
Delaunay triangulation, is equivalent to sorting the $x_i$
and thus as a $\Omega(n\log n)$ lower bound.

\section{Conclusion}

We have proposed a simple method for point deletion in Delaunay triangulations.
This method guarantees an $O(k\log k)$ complexity where $k$ is
the degree of the removed point while most  alternatives
have a quadratic behavior in the worst case.
The implementation  in two dimensions corroborates these results and
shows a good behavior in practice.
The algorithm should be even more efficient in higher dimensions
due to the lack of alternative methods and the higher average degree
of a Delaunay vertex.

\newpage

{\bf Code}
A  compiled demo version
is available at \\
http://www.inria.fr/prisme/logiciels/del-hierarchy/.

{\bf Acknowledgment}
The author would like to thank Jean-Michel Moreau,
Jack Snoeyink  and Mariette Yvinec
for helpful discussions and careful reading of this paper.

\newpage

\bibliography{geom,prisme,local}
\bibliographystyle{alpha}

\end{document}